\documentclass[a4paper,11pt]{article}
\pdfoutput=1 
\usepackage{jheppub} 
\usepackage{color}
\usepackage{graphicx, multirow,soul,amsmath,amsfonts,amssymb,mathrsfs,amsfonts}
\usepackage{hyperref}
\usepackage[all]{hypcap}
\usepackage{url}
\usepackage{bbm}
\usepackage{cancel}
\usepackage{slashed}
\usepackage[dvipsnames]{xcolor}
\usepackage{multicol, blindtext}
\usepackage{lipsum}
\usepackage{mattens}
\usepackage{mathtools}
\usepackage{footnote}
\usepackage{subcaption}
\usepackage[utf8]{inputenc}
\usepackage{cleveref}

\newcommand{\mathematica}{\textsc{Mathematica}}
\newcommand{\pythia}[1]{\textsc{Pythia#1}}

\newcommand{\madgraph}[1]{\textsc{MadGraph#1}}

\newcommand{\feynrules}[1]{\textsc{FeynRules#1}}

\newcommand{\delphes}[1]{\textsc{Delphes#1}}

\newcommand{\met}{\slashed{E}_{\mathrm{T}}}

\newcommand{\be}{\begin{equation}}
\newcommand{\ee}{\end{equation}}

\newcommand{\Mbivo}{\ensuremath{M_{\tilde{B}}}}

\newcommand{\bivo}{bi$\nu$o~}
\newcommand{\Bivo}{Bi$\nu$o~}

\preprint{\\ UCI-HEP-TR-2021-07}

\title{Long-lived bi$\boldsymbol{\nu}$o at the LHC}
\author[a]{Julia Gehrlein,}
\author[b]{Seyda Ipek}

\affiliation[a]{High Energy Theory Group, Physics Department, Brookhaven National Laboratory, Upton, NY 11973, USA}
\affiliation[b]{Department of Physics and Astronomy, University of California, Irvine
4129 Frederick Reines Hall, Irvine, CA 92617-4575, U.S.A.}

\abstract{We examine the detection prospects for a  long-lived bi$\nu$o, a pseudo-Dirac bino which is responsible for neutrino masses, at the LHC and at dedicated long-lived particle detectors. The \bivo arises in $U(1)_R$-symmetric supersymmetric models where the neutrino masses are generated through higher dimensional operators in an inverse seesaw mechanism. 
At the LHC the \bivo is produced through squark decays and it subsequently decays to quarks, charged leptons and missing  energy  via  its  mixing  with  the  Standard  Model  neutrinos.  We consider long-lived bi$\nu$os which escape the ATLAS or CMS detectors as missing energy and decay to charged leptons inside the proposed long-lived particle detectors FASER, CODEX-b, and MATHUSLA. We find the currently allowed region in the squark-\bivo mass parameter space by recasting most recent LHC searches for jets+$\slashed{E}_T$. We also determine the reach of MATHUSLA, CODEX-b and FASER. We find that a large region of parameter space involving squark masses, \bivo mass and the messenger scale can be probed with MATHUSLA, ranging from \bivo masses of 10 GeV-2 TeV and messenger scales $10^{2-11}$ TeV for a range of squark masses.}

\begin{document} 
\maketitle
\flushbottom

%%%%%%%%%%%%%%%%%
\section{Introduction}
\label{sec:intro}

There are several open questions in particle physics which cannot be answered by the Standard Model (SM). Amongst the most pressing ones are the generation of neutrino masses, the need for a dark matter candidate, and a mechanism to generate the observed baryon asymmetry of the universe. As searches for new physics particles in different environments have so far come up empty handed, novel search strategies need to be developed. 

One possibility is to extend the searches to look for particles which do not decay promptly at particle colliders but have a macroscopic decay length. In the last several years large interest has arisen to search for long-lived particles (LLPs) as several detectors at the LHC have been proposed  \cite{Chou:2016lxi, Alpigiani:2018fgd,Alpigiani:2020tva,Feng:2017uoz, Ariga:2018zuc,Ariga:2019ufm, Gligorov:2017nwh, Aielli:2019ivi, Gligorov:2018vkc} and detailed investigations from the model building side have been done (see \cite{Alimena:2019zri, Lee:2018pag} for recent reviews). Many minimal supersymmetric models (MSSM) naturally give rise to LLPs such that they can serve as benchmark models for various analyses. In most supersymmetric models the LLP is the next-to-lightest supersymmetric particle (NLSP) which decays into the lightest supersymmetric particle. But, for example, $R-$parity violating supersymmetry allows a final state which contains purely SM particles.

In this work we study the prospect of probing the parameter space of a certain $R$-symmetric MSSM model, introduced in \cite{Coloma:2016vod} and studied in more detail in \cite{Fox:2019ube}, via LLP searches. In $R-$symmetric MSSM the superpartners are charged under a global $U(1)_R$ symmetry while the SM particles are neutral \cite{Hall:1990hq}. Due to this global symmetry gauginos are expected to be Dirac fermions. However the global $U(1)_R$ is broken because the gravitino acquires a mass, which then leads to small $U(1)_R$--breaking Majorana masses for gauginos such that they are pseudo-Dirac fermions having both Dirac and Majorana masses \cite{Randall:1998uk,Giudice:1998xp,ArkaniHamed:2004yi}.\footnote{To obtain Dirac masses for the gauginos additional adjoint fields with opposite $U(1)_R$ need to be introduced \cite{Fayet:1974pd,Fayet:1975yi}. See \Cref{sec:model}.}
The collider phenomenology of $R-$symmetric MSSM differs from the MSSM phenomenology. 
For example, some production channels for supersymmetric particles are not available due to the $U(1)_R$ symmetry. In general collider limits on $R-$symmetric MSSM tend to be less stringent than the ones on MSSM \cite{Frugiuele:2012kp,Alvarado:2018rfl,Diessner:2017ske,Kalinowski:2015eca,Fox:2019ube}.

In the model we study, the $U(1)_R$ symmetry is elevated to $U(1)_{R-L}$, where $L$ is lepton number. It has been shown in \cite{Coloma:2016vod} that  the pseudo-Dirac bino, dubbed bi$\nu$o, in this model can play the role of right-handed neutrinos and that light Majorana neutrino masses are generated via an inverse-seesaw mechanism. The smallness of the light neutrino masses is generated by a hierarchy between the source of $U(1)_R$ breaking, namely the gravitino mass $m_{3/2}$, and the messenger scale $\Lambda_M$. Furthermore, the decay rate of the \bivo is inversely proportional to the messenger scale. Hence probing different possible lifetimes of the \bivo provides valuable information on the messenger scale and the origin of neutrino masses in this model. 

In \cite{Fox:2019ube} \bivo decays at the LHC were investigated in order to find the constraints on squark and \bivo masses. In that work the messenger scale was set to $\Lambda_M=100~$TeV, a scale expected to be probed by low energy experiments like Mu2e, and only \bivo masses $O(100~{\rm GeV})$ were investigated. At this scale the \bivo decays promptly and the strongest constraints come from ATLAS jets+$\slashed{E}_T$ search, where the missing energy comes from neutrinos produced in \bivo decays in contrast to the LSP as in most other MSSM models. As the messenger scale rises, \bivo decay width becomes smaller, making the \bivo long-lived at LHC scales. In this work we study the prospects of probing $\Lambda_M > 10^5~$TeV. At these scales \bivo would be considered as missing energy in ATLAS and CMS jets+$\slashed{E}_T$ searches, while it can decay into charged leptons inside proposed LLP detectors like MATHUSLA, FASER and CODEX-b. To this aim we will contrast the constraints from jets+$\slashed{E}_T$ searches at the LHC with $\sqrt{s}=13$ TeV and $\mathcal{L}=36~\text{fb}^{-1}$ with forecasted searches using the same channel with  $\mathcal{L}=3~\text{ab}^{-1}$ at MATHUSLA, FASER and CODEX-b. We  show that while FASER and its upgraded version FASER 2 are not competitive against jets+$\slashed{E}_T$ searches, MATHUSLA and CODEX-b can probe the messenger scale over a wide range of parameter space, $\Lambda_M \sim 10^{5-12}~$TeV for \bivo and squark masses of $O(100~{\rm GeV}-{\rm TeV})$. Our results are given in \Cref{fig:LLPexc}.

This manuscript is organized as follows: we give a brief overview of the model in \Cref{sec:model}, in \Cref{sec:bivolifetime} we provide analytical results for the \bivo lifetime and decay length, \Cref{sec:LHCpheno} is devoted to our numerical study of the bino phenomenology including the reach of LHC and LLP searches, and we summarize and conclude in \Cref{sec:summary}.

%%%%%%%%%%%%%%%%%%%%%%%%%%%
\section{Model}
\label{sec:model}

\begin{table}
\centering
\begin{tabular}{|c|c|c|}
\hline
Superfields	&	$U(1)_R$	&	$U(1)_{R-L}$ \\
\hline\hline
$L$	&	1	&	0 \\
$E^c$	&	1	&	2	\\	
\hline
$H_{u,d}$	&	0	&	0	\\
$R_{u,d}$	&	2	&	2	\\
\hline
$W_{\tilde{B}}$	&	1	&	1	\\
$\Phi_{S}$	&	0	&	0	\\
\hline
gravitino/goldstini & 1 &  1 \\ 
\hline
\end{tabular}
\caption{The relevant field content of the $U(1)_R$ symmetric model (SM charges not shown). $L$, $E^c$ are the lepton superfields and $H_{u,d}$ are the up-type and down-type Higgs superfields. The fermionic components of the superfields $R_{u,d}$ are the Dirac partners of the Higgsinos $\tilde{h}_{u,d}$. $\Phi_S$ is a  superfield which has the same SM charges as $W_{\tilde{B}}$ and its fermionic component $S$ is the Dirac partner of the bino.} \label{table:fields}
\end{table}

The details of this model have been given in \cite{Coloma:2016vod, Fox:2019ube}. In this section we briefly summarize the salient points important for our analysis. 

The model we work with is a modified version of $U(1)_R-$symmetric MSSM, in which we impose a global $U(1)_{R-L}$ symmetry, where $L$ is the lepton number, on the supersymmetric sector. SM particles are not charged under $U(1)_R$, but of course some of them have lepton charges. Employing the lepton number is essential to generating the interactions between bino and the SM neutrinos. Some of the supersymmetric fields and their $U(1)_R, U(1)_{R-L}$ charges are given in \Cref{table:fields}.

This model inherits many properties of $U(1)_R-$symmetric MSSM. The most prominent feature of these models is that gauginos are Dirac due the global $U(1)$ symmetry. As such, for each gaugino with $U(1)_R$ charge of 1, a Dirac partner is introduced with $-1$ $R$ charge. Here we will only focus on the bino, $\tilde{B}$, and its Dirac partner the singlino, $S$. 

Supersymmetry is broken in a hidden sector which communicates with the visible sector at a messenger scale $\Lambda_M$ and is incorporated via $F-$ and $D-$term spurions, $X=\theta^2 F$ and $W'_\alpha=\theta_\alpha D$ respectively. The $F-$term generates masses for the sfermions while the $D-$term spurion generates the Dirac gaugino masses via the supersoft term~\cite{Fox:2002bu}
\begin{align}
\int d^2\theta \frac{\sqrt{2}c_i}{\Lambda_M} W'_\alpha W_{\tilde{B}}^\alpha \Phi_S~,
\end{align}
 where $\Phi_S$ is the superfield whose fermionic component is the singlino $S$. The Dirac bino mass is $\Mbivo = c_i D/\Lambda_M$. 
 
The global $U(1)_{R-L}$ symmetry, as all global symmetries, is broken due to gravity. Hence a Majorana mass for the bino is generated via anomaly mediation \cite{Randall:1998uk,Giudice:1998xp,ArkaniHamed:2004yi}
\begin{align}
m_{\tilde{B}} = \frac{\beta(g_Y)}{g_Y} f_\phi~,
\end{align}
where $\beta(g_Y)$ is the beta function for the hypercharge and $F_\phi$ is a conformal parameter satisfying
\begin{align}
\frac{m_{3/2}^3}{16\pi^2 M_{\rm Pl}^2}<f_\phi<m_{3/2}~. 
\end{align}
$m_{3/2}^2=\sum (F_i^2 + D_i^2/2)/\sqrt{3}M_{Pl}^2$ is the gravitino mass. A Majorana mass for the singlino, $m_S$, is also expected to be produced, as well as Majorana masses for all other gauginos and their Dirac partners. We assume the messenger scale $\Lambda_M$ is below the Planck scale and $m_{\tilde{B}}, m_S \ll \Mbivo$. In the following we use the term \bivo to refer to the pseudo-Dirac fermion $\Psi_{\tilde{B}}^T = (\tilde{B},S^\dagger)$ and its Weyl component $\tilde{B}$ interchangeably, which should be clear from the context. We use $M_{\tilde{B}}$ as the \bivo mass, even though it gets small corrections from the Majorana masses as well.

It has been shown in \cite{Coloma:2016vod} that the operators,\footnote{These operators can be generated by integrating out two pairs of gauge singlets $N_i,N_i'$, with R-charge 1 and lepton number $\mp 1$.}
\be
\frac{f_i}{\Lambda_M^2}\int d^2\theta\, W'_\alpha W_{\tilde{B}}^\alpha H_u L_i \ \ \ \text{and}\ \ \ \frac{d_i}{\Lambda_M}\int d^4\theta\,  \phi^\dagger \Phi_S H_u L_i
\ee
(where $\phi=1+\theta^2 m_{3/2}$) can generate two non-zero neutrino masses through the inverse seesaw mechanism
\cite{Mohapatra:1986aw,Mohapatra:1986bd}, with the bino--singlino pair acting as a pseudo-Dirac right-handed neutrino.  Written in terms of their component fields, the neutrino-mass part of the Lagrangian becomes
\begin{align}
\mathcal{L}\supset M_{\tilde{B}} \tilde{B}S+m_{\tilde{B}}\tilde{B}\tilde{B}+m_S\, SS+ f_i \frac{M_{\tilde{B}}}{\Lambda_M} \ell_i h_u \tilde{B} + d_i \frac{m_{3/2}}{\Lambda_M} \ell_i h_u S~, 
\end{align}
where $f_i$ and  $d_i$, for $i=e,\mu,\tau$, are determined by neutrino mass differences as
\begin{align}
f_i \simeq 
\begin{pmatrix}
0.35 \\
0.85 \\
0.35
\end{pmatrix},\quad
d_i\simeq 
\begin{pmatrix}
-0.06 \\
0.44 \\
0.89
\end{pmatrix}~. \label{eq:YG}
\end{align}
After electroweak symmetry breaking the light neutrino masses are 
\begin{align}
m_1 = 0,~~~m_2=\frac{m_{3/2}\, v^2}{\Lambda_M^2}(1-\rho),~~~ m_3=\frac{m_{3/2}\, v^2}{\Lambda_M^2}(1+\rho)~,
\end{align}
where $\rho\simeq 0.7$ is determined by the neutrino mass splittings. 

\begin{figure}[t]
\centering
\includegraphics[width=.6\textwidth]{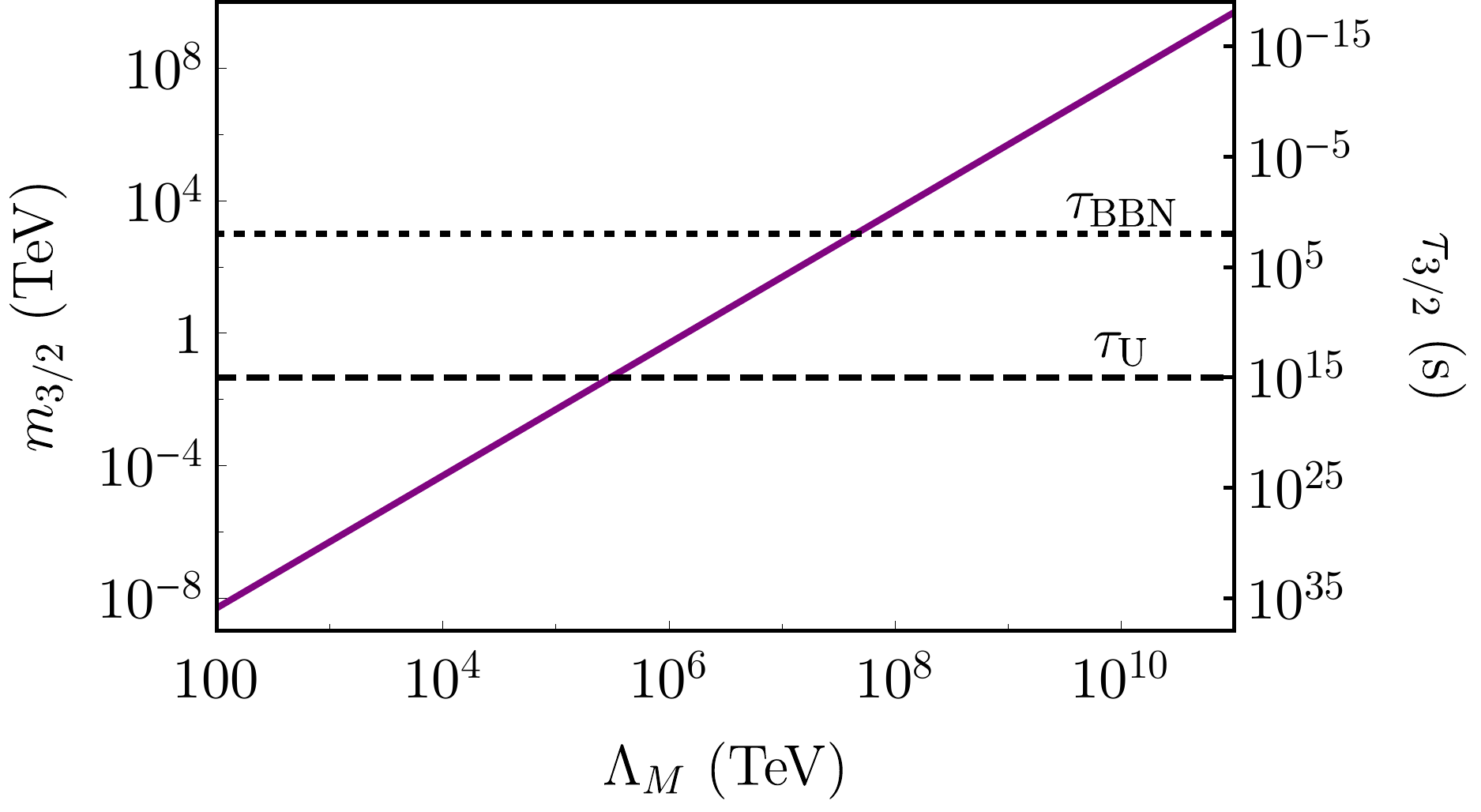}
\caption{Gravitino mass required to explain neutrino masses as a function of messenger scale $\Lambda_M$ for normal ordering. On the right y-axis we show the corresponding gravitino lifetime. The dashed horizontal line is the lifetime of the universe, $\tau_{\rm U}$, and the dotted line shows the time of BBN, $\tau_{\rm BBN}$, as reference points.}\label{fig:m32}
\end{figure}

In the following we will study the current and future LHC constraints and forecasted constraints from future LLP experiments on the parameter space of this model. It will be shown that for a long-lived \bivo which is observable at the LHC, the messenger scale needs to be high, up to $10^{11}$ TeV. This then requires a heavier gravitino than what was considered in earlier work. In \Cref{fig:m32} we show the gravitino mass as a function of $\Lambda_M$ for normal hierarchy. (Inverted hierarchy is very similar.) For $\Lambda_M\gtrsim 10^5$ TeV, the gravitino needs to be heavier than $\sim$TeV in order to explain the neutrino masses.\footnote{In order to have both LHC-accessible squarks and a heavy gravitino, we imagine a scenario like \cite{Fox:2019ube} where there are two SUSY-breaking scales, with  $\sqrt{D_1},\sqrt{F_1}\sim 10$ TeV and $\sqrt{D_2},\sqrt{F_2}\gtrsim 10^4$ TeV.} The gravitino in this model decays primarily into photons and neutrinos with decay width $\Gamma_{3/2}\simeq \theta^2 m_{3/2}^3/M_{\rm Pl}^2$, where $\theta\sim 10^{-3}$ is the neutrino-\bivo mixing angle and $M_{\rm Pl}\simeq 1.2\times 10^{16}$ TeV is the Planck mass. In \Cref{fig:m32} we also show the gravitino lifetime. In earlier work~\cite{Fox:2019ube} it was mentioned that a gravitino with $O(10~{\rm keV})$ could be a dark matter candidate since it is stable within the lifetime of the universe. As the gravitino gets heavier, this decay width becomes larger.  Above $m_{3/2}\sim O(10~{\rm GeV})$ the gravitino is no longer stable enough to constitute the dark matter. If the gravitino decays around the time of Big Bang Nucleosynthesis (BBN), the produced photons could affect the production of light nuclei. In our model this could happen for $m_{3/2}\sim O(10^{1-3}~{\rm TeV})$, corresponding to $\Lambda_M\sim 10^7$ TeV. Note that there could also be new decay channels opening up if the gravitino is no longer the LSP, affecting the decay width. A detailed study of the gravitino is beyond the scope of this work. For now we ignore details of the gravitino behavior in this wide mass range and assume there are mechanisms, \emph{e.g.} a low reheat temperature, to suppress the abundance of gravitinos with certain mass so that BBN proceeds as observed.

%%%%%%%%%%%%%%%%%%%%%%%%%%%
\section{Bi$\boldsymbol{\nu}$o lifetime}
\label{sec:bivolifetime}
%%%%%%%%%%%%%%%%%%%%%%%%%%%

The \bivo decays primarily via its mixing with light neutrinos, with branching fraction 1/3 to each channel: (i) $\tilde{B}\to W^-\ell^+$; (ii) $\tilde{B}\to Z\bar{\nu}$; and (iii) $\tilde{B}\to h\bar{\nu}$. 
Note that the decays to gravitinos is strongly suppressed by the Planck mass, $\Gamma(\tilde{B}\to\tilde{G}\gamma)\sim \frac{M_{\tilde{B}}^5}{M_{\rm Pl}^2 m_{3/2}^2}\sim 10^{-8}$~eV, and  the $\tilde{B}\to W^+\ell^-$ decay is not allowed due to the $U(1)_{R-L}$ symmetry. 

There are two interesting mass regimes for the bi$\nu$o: \textbf{(i)} $\Mbivo > M_{Z,W,h}$; and \textbf{(ii)} $\Mbivo < M_{Z,W,h}$, which we will refer to as heavy and light \bivo respectively. For simplicity we will consider $\Mbivo > 125~$GeV and $\Mbivo < 80~$GeV for each regime. 

\textbf{Heavy bi$\boldsymbol{\nu}$o} In this scenario the \bivo can decay into on-shell $W,Z,h$ final states via a 2-body process. The total decay width is
\begin{align}
\Gamma_{\tilde{B}}^{heavy}\simeq \sum_{i=e,\mu,\tau} \Mbivo Y_i^2 \simeq \frac{\Mbivo^3}{\Lambda_M^2}~, ~~~~{\rm where}~~ Y_i=f_i\frac{\Mbivo}{\Lambda_M}~.
\end{align}

\textbf{Light bi$\boldsymbol{\nu}$o} In this regime, since the \bivo is lighter than the gauge bosons, it has to decay through off-shell $W,Z,h$ to 3-body final states. 

\begin{align}
\Gamma_{\tilde{B}}^{light} \simeq \sum_{i=e,\mu,\tau}  \kappa\, Y_i^2\frac{G_F^2 \Mbivo^5}{192\pi^3}\simeq \kappa \frac{G_F^2 \Mbivo^7}{192\pi^3\Lambda_M^2},
\end{align}
where $\kappa$ is an $O(1)$ number which encodes phase-space integrals. For our analysis we will take $\kappa = 1$.

\begin{figure*}[t]
  \centering
    \begin{subfigure}[b]{0.525\textwidth}
  \includegraphics[width=\textwidth]{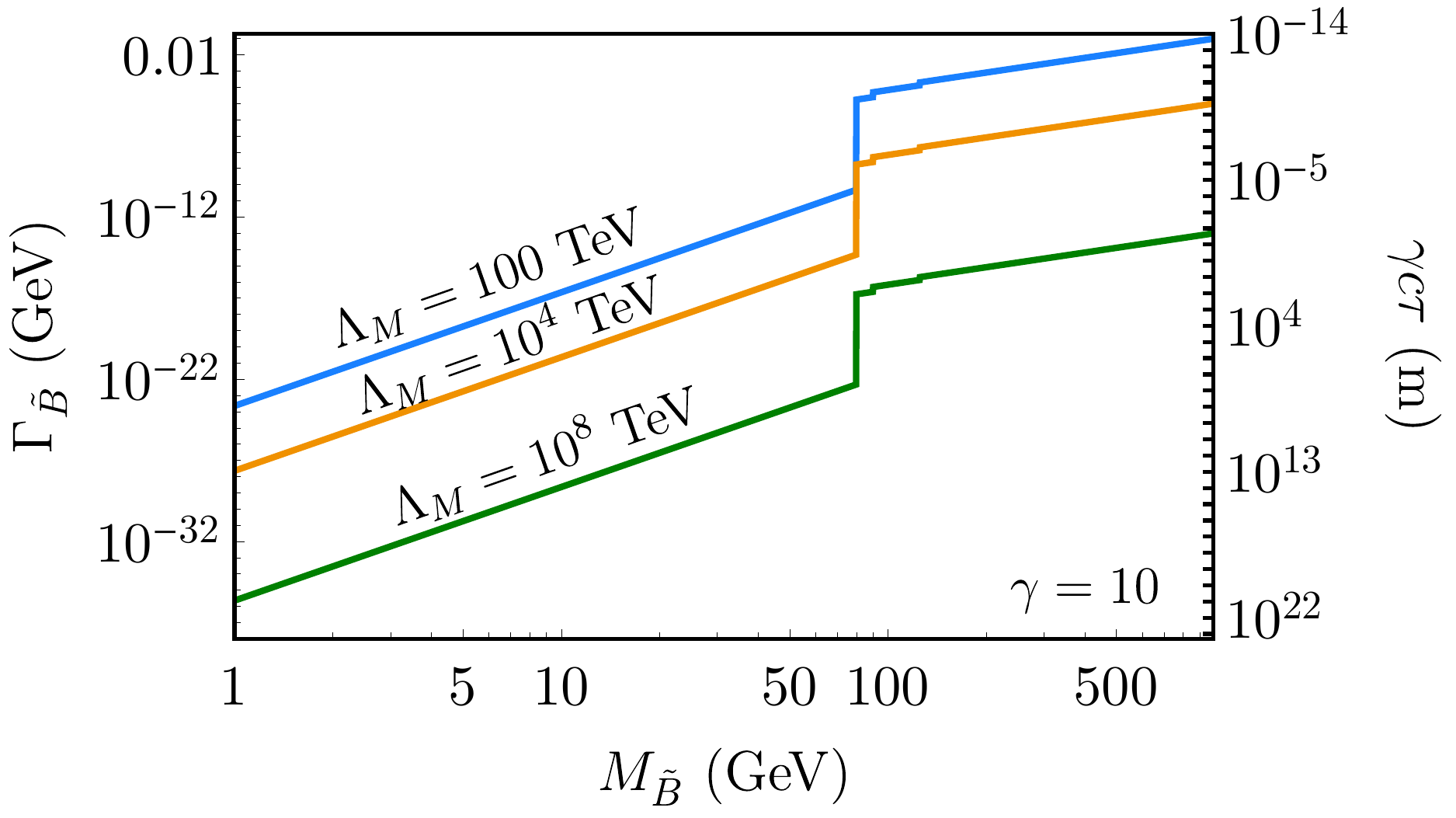}\label{fig:Gamma}
  \caption{}
  \end{subfigure}%
    ~~~~~
  \begin{subfigure}[b]{0.47\textwidth}
 \includegraphics[width=\textwidth]{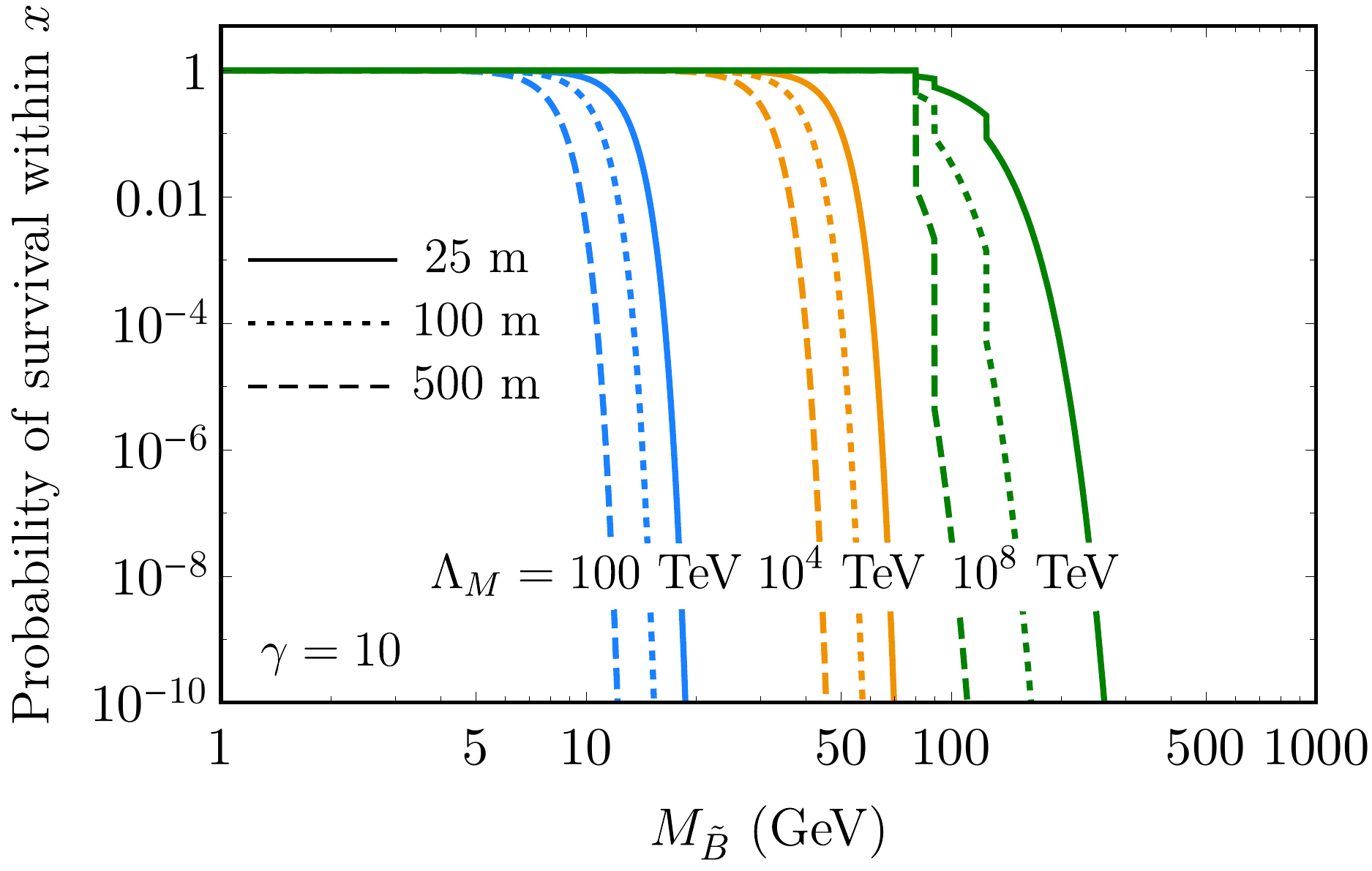}\label{fig:probdecay}
 \caption{}
 \end{subfigure}
  \caption{\textbf{(a)} \Bivo decay width and decay length versus the \bivo mass \Mbivo. We show three different choices of the messenger scale, $\Lambda_M= 10^2, 10^4, 10^8~$TeV. Below $\Mbivo=80~$GeV \bivo decays to 3-body final states through off-shell gauge and Higgs bosons and hence the decay width is much smaller. \textbf{(b)} Probability of a \bivo surviving, i.e. \emph{not} having decayed, after traveling a distance $x=25,100,500~$ meters, roughly corresponding to the length of the ATLAS detector, and the distance between the \bivo production point and MATHUSLA and FASER respectively. The Lorentz factor is taken to be $\gamma = 10$, which corresponds to a momentum of $\sim $ TeV for a 100 GeV bi$\nu$o, as a benchmark value.}  \label{fig:decays}
\end{figure*}

On the left panel in \Cref{fig:decays} we show the \bivo lifetime for various messenger scales. It can be seen that the width drops several orders of magnitude below the $\Mbivo\sim M_W$ threshold.  On the same plot, we also give the decay length $\gamma c\tau$ of \bivo assuming a benchmark boost factor of $\gamma=10$, which shows the expected behavior where as the decay width gets smaller, the lifetime gets longer. It can be seen in this plot that for $\Lambda_M=100$ TeV and $\Mbivo>100$ GeV, which was covered in \cite{Fox:2019ube}, \bivo decays within a nanometer of the production point. However for lighter \bivo and larger messenger scale, \bivo lifetime can be much longer. 

For the long-lived particle searches, we are interested in how far the \bivo travels before it decays. Interesting length scales are related to the size and placement of certain detectors. For example the ATLAS detector is $\sim 25~$meters in diameter while MATHUSLA is planned to be placed $\sim 70$ meters from CMS~\cite{Alpigiani:2020tva} and FASER is $\sim 500$ meters away from ATLAS interaction points~\cite{Ariga:2019ufm}. The probability that a relativistic particle survives at a distance $x$ away from its production point is
\begin{align}
P(x)=\exp(-x/\gamma  c\tau)~,
\end{align}
where $\gamma = 1/\sqrt{1-v^2/c^2}$ is the Lorentz factor and $\tau = 1/\Gamma$ is the lifetime of the particle at its rest frame. In the right panel of \Cref{fig:decays} we show the probability of survival of a \bivo at various distances from the production point as a function of the \bivo mass. For low messenger scales, \emph{e.g.} for $\Lambda_M = 100$ TeV, the probability that the \bivo will travel hundreds of meters is low for $\Mbivo \gtrsim 10 $ GeV. (Note that messenger scales below 100 TeV can be probed at the upcoming Mu2e experiment~\cite{Bartoszek:2014mya,Coloma:2016vod} independent of the \bivo mass.) As the messenger scale gets larger, heavier bi$\nu$os could survive longer distances with much higher probability. 

The survival probability depends on the \bivo momentum, in terms of the Lorentz factor $\gamma$. In this work we will be looking at \bivo production from squark decays. Hence the energy and the velocity of the \bivo depends on the squark energy and momentum as well as the \bivo mass itself. Furthermore we will require the \bivo to decay within a certain detector volume that is placed at a certain angle from the interaction point. Hence the numerical calculation of the probability factor is more involved. We describe the experiments we consider and our analysis in the next section. 

%%%%%%%%%%%%%%%%%%%%%%%%%%%%%
\section{Long-lived bi$\boldsymbol{\nu}$o at the LHC}
\label{sec:LHCpheno}
In earlier work \cite{Fox:2019ube} the parameter space for a short-lived \bivo -- with proper decay length $c\tau \lesssim100~\mu$m -- was investigated. It has been found that current constraints go up to only $M_{\rm sq}\approx$ 950 GeV and squarks as light as 350 GeV are allowed for $M_{\tilde{B}}= 100-150$ GeV. In the previous section we showed that  \bivo can be long lived if its mass is low and/or if the messenger scale is high. In this section we examine search strategies to look for a long-lived \bivo at the LHC. Our results are shown in \Cref{fig:LLPexc}.

\begin{figure}
\centering
\includegraphics[width=.8\textwidth]{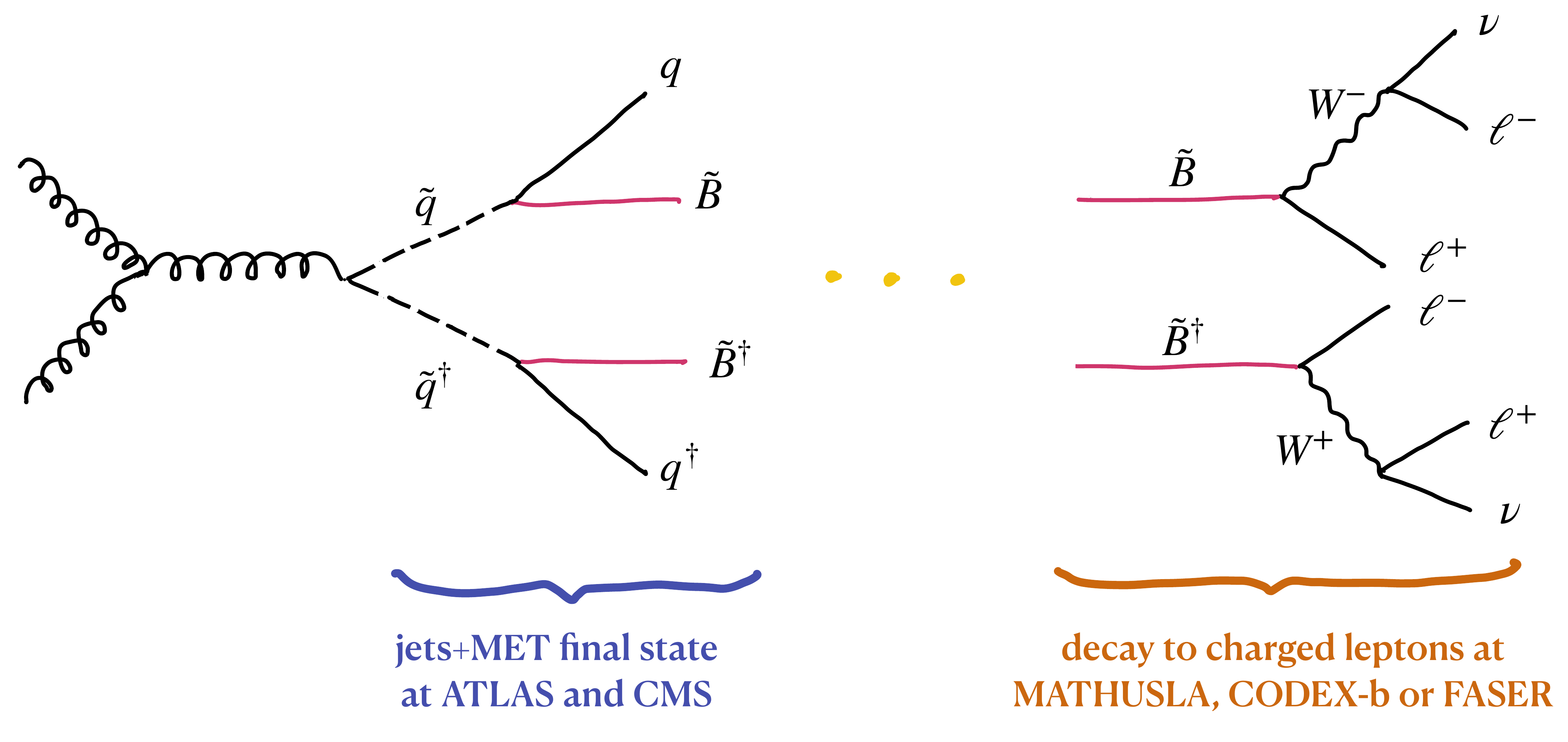}
\caption{An example process for \bivo production via squark decays and subsequent decay of a long-lived \bivo to charged leptons.}\label{fig:LLPsignal}
\end{figure}

\begin{figure*}[t]
  \centering
    \begin{subfigure}[b]{0.5\textwidth}
  \includegraphics[width=\textwidth]{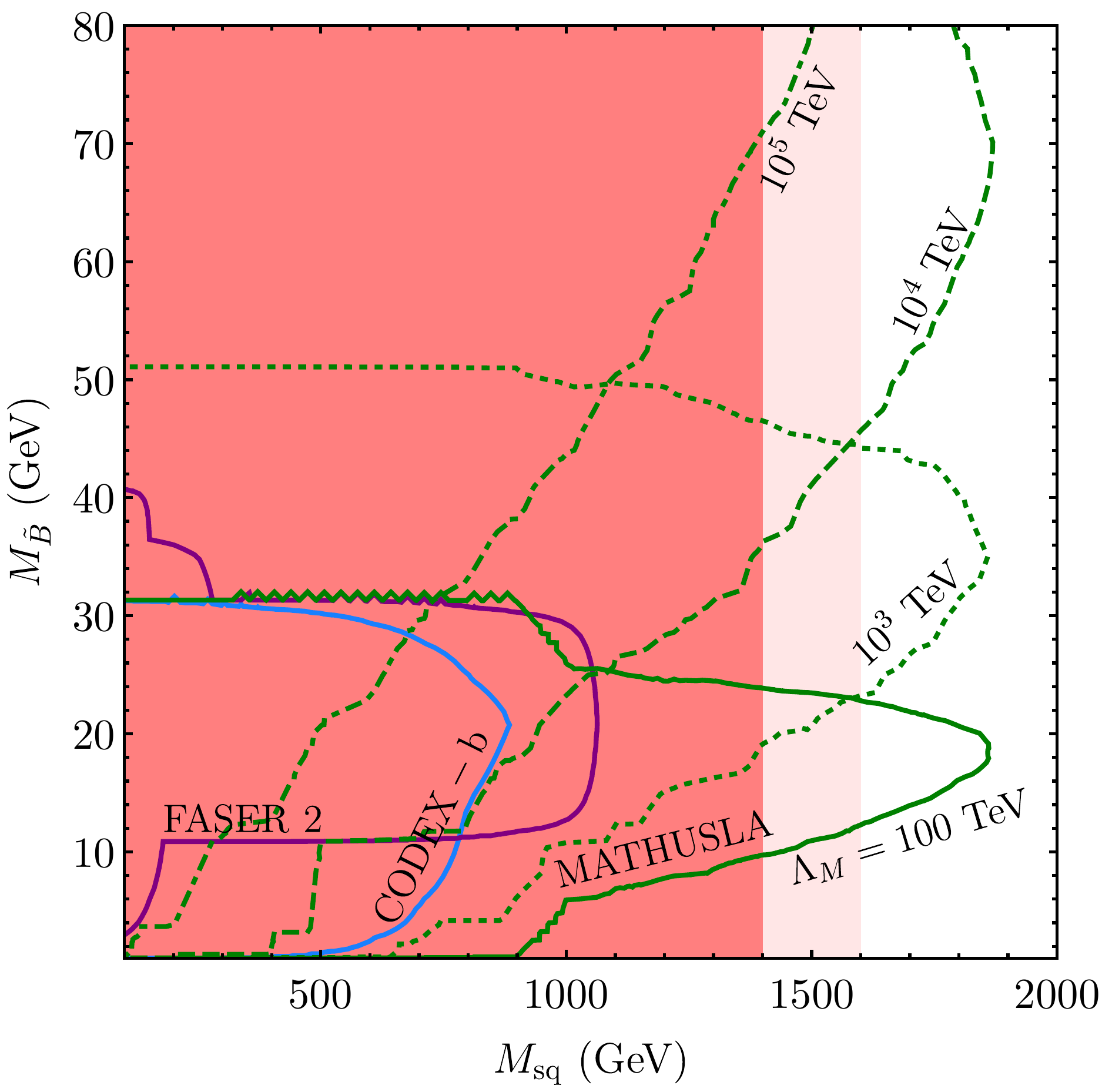}\label{fig:lowMBexc}
  \caption{}
  \end{subfigure}%
    ~~~
  \begin{subfigure}[b]{0.52\textwidth}
 \includegraphics[width=\textwidth]{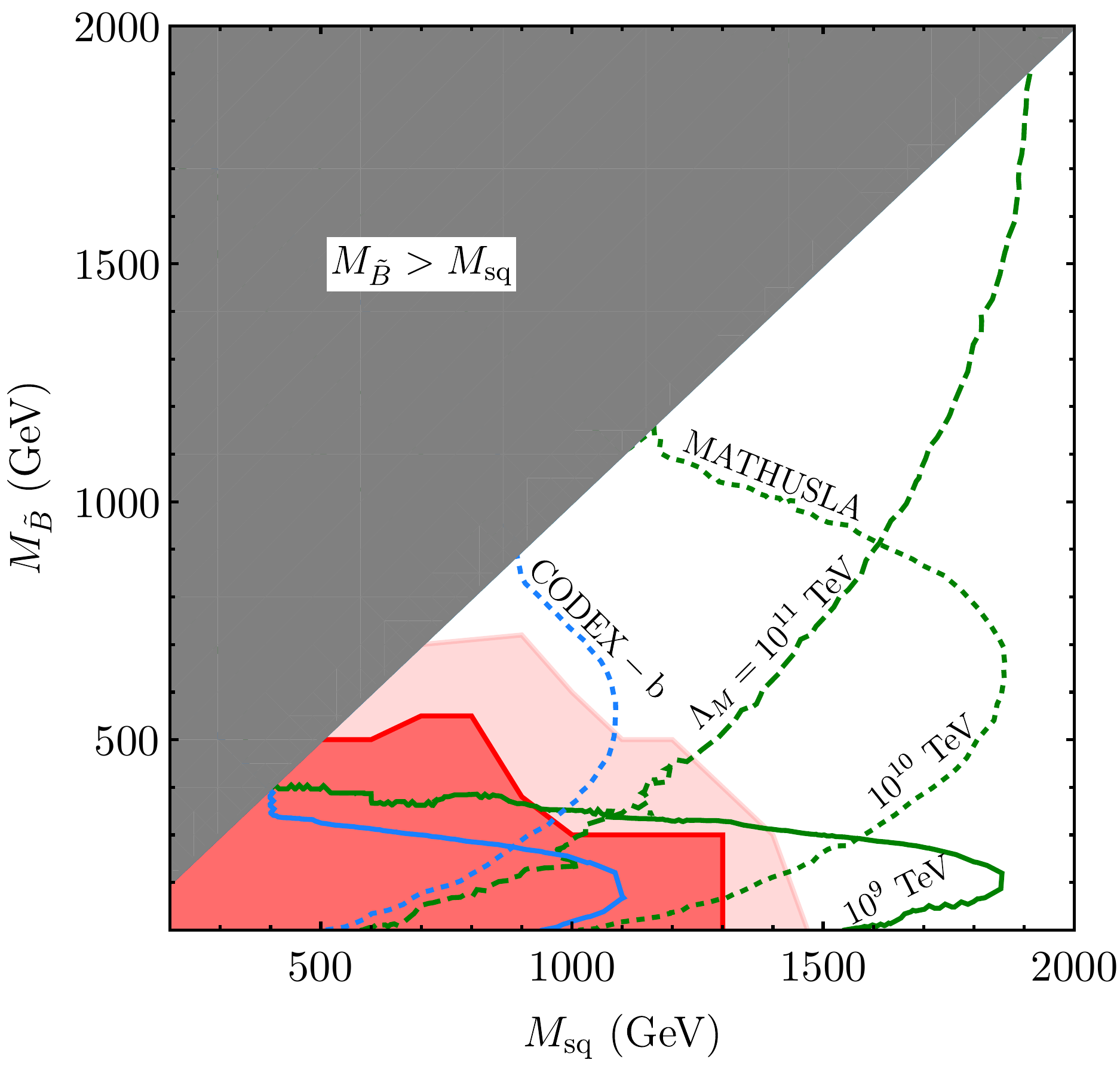}\label{fig:highMBexc}
 \caption{}
 \end{subfigure}
  \caption{\textbf{(a)} Current and future constraints on light \bivo mass region. Dark red shaded region shows the LHC constraint with $\mathcal{L}=36~{\rm fb}^{-1}$ of data while light pink is projection to $\mathcal{L}=3~{\rm ab}^{-1}$. The colored lines correspond to 5 events in each detector labeled with $\mathcal{L}=3~{\rm ab}^{-1}$. We show lines of different messenger scales, $\Lambda_M = 10^{2-5}$ TeV, for MATHUSLA. We show only $\Lambda_M=100$ TeV for FASER 2 and CODEX-b because the parameter space that can be reached by these two experiments is already ruled out by ATLAS jets+$\slashed{E}_T$ search. \textbf{(b)} Current and future constraints on heavy \bivo mass region. Dark red shaded region shows the LHC constraint with $\mathcal{L}=36~{\rm fb}^{-1}$ of data while light pink is projection to $\mathcal{L}=3~{\rm ab}^{-1}$. The green and blue colored lines correspond to 5 events in MATHUSLA and CODEX-b respectively with $\mathcal{L}=3~{\rm ab}^{-1}$. We show lines of different messenger scales, $\Lambda_M = 10^{9-11}$ TeV. } \label{fig:LLPexc}
\end{figure*}

We assume a sparticle spectrum where first- and second-generation squarks are degenerate and third generation is decoupled. We assume sleptons, gluinos and charginos are also heavy and decoupled. The lightest neutralino is a pure \bivo and the other neutralinos are heavier than the degenerate squarks. (See Fig.1 of \cite{Fox:2019ube}.) Depending on the gravitino mass, \bivo is either the LSP or the NLSP. Due to the sparticle spectrum we assume, \bivo is produced mainly through squark decays. After being produced the long-lived \bivo decays into a combination of quarks, charged leptons and neutrinos. (See \Cref{fig:LLPsignal} for an example process.) Such an event gives two complimentary search possibilities at the LHC: (i) a long-lived \bivo  escapes the LHC as missing energy so that it can be searched for by the jets+$\slashed{E}_T$ searches; and  (ii) \bivo decays to charged leptons inside an LLP detector.
There are several (proposed or recently being built) dedicated LLP detectors like MATHUSLA \cite{Chou:2016lxi, Alpigiani:2018fgd,Alpigiani:2020tva},
FASER \cite{Feng:2017uoz, Ariga:2018zuc,Ariga:2019ufm}, and CODEX-b \cite{Gligorov:2017nwh, Aielli:2019ivi}.
Additionally, future ATLAS or CMS LLP searches can be sensitive to long-lived \bivo decays \cite{ATL-PHYS-PUB-2019-013, ATL-PHYS-PUB-2018-033}. Here we will focus on the reach of proposed, dedicated LLP experiments.

%%%%%%%%%%%%%%%%%%%%%%
\subsection{Missing energy analysis}
\label{sec:signals}
%%%%%%%%%%%%%%%%%%%%%%
A long-lived \bivo with $c\tau\gtrsim 100~ \mu$m will leave the inner tracker system of ATLAS as missing energy such that missing energy searches can be used constrain the $M_{\rm sq}-\Mbivo$ parameter space. The signature at the LHC is a 2j (from the \bivo production) + $\slashed{E}_T$ final state for which we can recast the multijet+$\slashed{E}_T$ ATLAS analysis \cite{ATLAS-CONF-2017-022}. 
We implement our model in \feynrules{}
 \cite{Alloul:2013bka} and generate 10K events with \madgraph{5} \cite{Alwall:2014hca}, using \pythia{8} \cite{Sjostrand:2014zea} for parton shower and hadronization, and \delphes{} \cite{deFavereau:2013fsa} for detector simulation at $\sqrt{s}=13$ TeV and $\mathcal{L}=36~\text{fb}^{-1}$. We use the default settings for jets in \madgraph{5} with $R=0.4, ~p_{Tj}>20$ GeV and $|\eta_j|<5$. 
We generate signal events for bi$\nu$os in the mass range $1~{\rm GeV}<M_{\tilde{B}}<M_{\rm sq}$ where we take first and second generation squarks to be degenerate with mass $M_{\rm sq}$.\footnote{We also looked at decoupling the second generation. In this case jets+$\slashed{E}_T$ constraints are milder due to the lowered production cross-section. However, expected number of \bivo events at LLP detectors are also smaller due to the same cross-section suppression.}
All other sparticle masses have been set to 10~TeV such that they are decoupled. We  generated events with 10 GeV and 50 GeV intervals for $\Mbivo$ and $M_{\rm sq}$ respectively and extrapolated our results for the intervening masses. 

The $m_{\rm eff}$-based  analysis by ATLAS \cite{ATLAS-CONF-2017-022} relies on  the observable $m_{\rm eff}$  defined as the scalar sum of the transverse momenta of the leading jets and missing energy, $\met$. Taken together with $\met$, $m_{\rm eff}$ strongly suppresses the multijet background. There are 24 signal regions in this analysis. These regions are first divided according to jet multiplicities (2-6 jets).
 Signal regions with the same jet multiplicity are further divided according to the values of $m_{\rm eff}$ and the $\met/m_{\text{eff}}$ or $\met/\sqrt{H_T}$ thresholds. 
 In each signal region, different thresholds are applied on jet momenta and pseudorapidities to reduce the SM background.
Cuts on the smallest  azimuthal  separation  between $\met$ and 
 the momenta of any of the reconstructed jets further reduce the multi-jet background. Two of the signal regions require two large radius jets and in all  signal regions the required  jet momentum $p_T>50~$GeV and missing energy $\met>250~$GeV. 
 The thresholds on the observables which characterize the signal regions have been chosen to target models with squark or gluino pair production and direct decay of squarks/gluinos or one-step decay of squark/gluino via an intermediate chargino or neutralino.
In order to identify the allowed parameter points we compare the  signal cross-section to the measured cross-section limits at  95$\%$ C.L. in all 24 signal regions using the code from \cite{Asadi:2017qon}. If the signal cross-section of a parameter point exceeds the measured  cross-section at 95$\%$ C.L. in at least one bin we take this parameter point to be ruled out.

We also analyze the expected exclusion limits at the end of  LHC Run 3 with $\sqrt{s}=13~$TeV and $\mathcal{L}=3~\text{ab}^{-1}$, by rescaling the expected number of signal and background events  with luminosity. In order to obtain the allowed parameter region at a high-luminosity LHC we use the median expected exclusion significance \cite{Kumar:2015tna}\footnote{See \cite{Bhattiprolu:2020mwi} for alternative expressions for the expected significance.}

\begin{align}
Z_{exc}=\Big[2\left(s-b \log\left(\frac{b+s+x}{2b}\right)-\frac{b^2}{\Delta_b^2}\log\left(\frac{b-s+x}{2b}\right)\right)-(b+s-x)(1+\tfrac{b}{\Delta_b^2})
\Big]^{1/2}~,
\end{align}
with
\begin{align}
x=\big[(s+b)^2-4sb\tfrac{\Delta_b^2}{(b+\Delta_b^2)}\big]^{1/2}~,
\end{align}
where $s$ is the signal, $b$ is background and $\Delta_b$ is the uncertainty on the background prediction. For a 95$\%$ C.L. median exclusion, we require $Z_{exc}>1.645$.  We assume, as a conservative estimate, that the relative background uncertainty after $3~\text{ab}^{-1}$ remains the same as it is now, as presented in \cite{ATLAS-CONF-2017-022}.  The estimate that $\Delta_b/b$ is constant could be improved upon, especially if the background is estimated from data in sidebands.

Our results are shown in \Cref{fig:LLPexc} for $\sqrt{s}=13$ TeV. We divide our analysis and results into low and high \bivo mass regions. On the left panel of \Cref{fig:LLPexc} we show the constraints on the low \bivo mass regime with $1~{\rm GeV} < \Mbivo \lesssim 80$ GeV. In this case the \bivo is long lived even with $\Lambda_M=100$ TeV and is seen as missing energy at ATLAS. In this parameter region squarks up to 1.4 TeV are excluded at $\mathcal{L}=36~\text{fb}^{-1}$, red/dark shaded area, and squark masses up to 1.6 TeV will be probed with $\mathcal{L}=3~\text{ab}^{-1}$, pink/light shaded area.\footnote{For decoupled second generation, the current constraint is 1.2 TeV and future reach will 1.4 TeV.} The constraints on the squark mass are independent of the \bivo mass. On the right panel of \Cref{fig:LLPexc} is the high \bivo-mass region with $80~{\rm GeV}< \Mbivo<M_{\rm sq}$, where \bivo is long lived. (Note that the exact messenger scale that corresponds to a long-lived \bivo depends on $\Mbivo$ and \bivo energy. Generally it requires $\Lambda_M\gtrsim 10^4$ TeV.) In this regime the constraints on the squark masses depend on the \bivo mass. With $\mathcal{L}=36~\text{fb}^{-1}$, squark masses up to 1300 TeV are excluded for $\Mbivo < 300$ GeV. This limit can go up to 1450 GeV with $\mathcal{L}=3~\text{ab}^{-1}$.\footnote{Corresponding current and future constraints for a decoupled second generation scenario is $M_{\rm sq}=$ 1 TeV and 1.2 TeV respectively. } On the other hand, squarks as light as 600 GeV are still allowed for heavier bi$\nu$o.

%%%%%%%%%%%%%%%%
\subsection{LLP detectors}
\label{sec:LLPdetectors}

In this section we study the cases where the \bivo decays inside one of the LLP detectors, namely MATHUSLA, CODEX-b and FASER. The parameter space for such decays depends on the \bivo mass and the messenger scale, as can be observed in \Cref{fig:decays}. For this analysis we follow \cite{Dercks:2018eua}, in which the required steps were very clearly laid out.

The number of \bivo events that can be observed in a detector $D=$ MATHUSLA, FASER or CODEX-b is 
\begin{align}
N_{\tilde{B}}^D = \mathcal{L}\times\sigma_{\tilde{B}}\times Br \times P(\tilde{B}\in D)~,
\end{align}
where $\sigma_{\tilde{B}}$ is the \bivo production cross-section at the LHC at $\sqrt{s}=13$ TeV and $Br\simeq 0.7$ is the branching fraction of \bivo to charged leptons~\cite{Fox:2019ube}. We give our results for $\mathcal{L}=3~{\rm ab}^{-1}$ for MATHUSLA and FASER, but only $300~{\rm fb}^{-1}$ for CODEX-b since LHCb receives a tenth of the luminosity of CMS and ATLAS.  

$P(\tilde{B}\in D)$ captures the decay probability of \bivo inside the detector and needs to be calculated for each detector geometry. (See Section III of \cite{Dercks:2018eua} for details.) In order to find this quantity we generate 10K parton level $gg\to q\bar{q} \tilde{B}\tilde{B}^\dagger$ events with \madgraph{5} for each squark and \bivo mass point described in the previous section. We analyze the results in \mathematica{} with a routine written by D. Curtin~\cite{lhereader}. The probability that a \bivo particle with certain momentum will decay inside a detector depends on where the detector is and the detector geometry. Without going into too much detail, the necessary information about each detector can be summarized as follows.
 
\begin{itemize}
\item \textbf{MATHUSLA} \cite{Chou:2016lxi, Alpigiani:2018fgd,Alpigiani:2020tva} is a proposed experiment that will be 60 m in horizontal  and 68 m in vertical distance from the CMS interaction point. Its size will be $100~{\rm m}\times 25~{\rm m} \times 25~{\rm m}$. 

\item \textbf{CODEX-b} \cite{Gligorov:2017nwh, Aielli:2019ivi} is proposed to be 25 m from LHCb interaction point. It will be a cube of $10~{\rm m}\times 10~{\rm m}\times 10~{\rm m} $.

\item \textbf{FASER} \cite{Feng:2017uoz, Ariga:2018zuc,Ariga:2019ufm} is being built 470 m from ATLAS inside the LHC tunnel. In its first version it is a cylinder with 10 cm radius and 1.5 m length. It is planned to be upgraded to a radius of 1 m and length of 5 m in the next LHC upgrade. We forecast our results for FASER 2, the upgraded version, to show that even this larger version will not be able to probe the parameter space we study.
\end{itemize}

In \Cref{fig:LLPexc} we show contours of five charged lepton events at each detector for $\mathcal{L}=3~{\rm ab}^{-1}$. For low mass \bivo (left panel in \Cref{fig:LLPexc}) the parameter space relevant for FASER and CODEX-b is completely ruled out by current ATLAS jets+$\slashed{E}_T$ searches. For this mass region MATHUSLA can still be complementary to and even competitive against future ATLAS missing energy searches for a messenger scale up to $\sim 10^4$ TeV. For heavier bi$\nu$o  (right panel in \Cref{fig:LLPexc}), with mass larger than $\sim 100$ GeV, the messenger scale that can be probed is higher. In this regime MATHUSLA will be able to detect events for $10^9~{\rm TeV}\lesssim\Lambda_M\lesssim10^{11}~{\rm TeV}$ with $3~{\rm ab}^{-1}$ luminosity. Most importantly, the squark mass that can be probed by MATHUSLA is much higher, $\sim 1.9$ TeV, than what will be probed by jets+$\slashed{E}_T$ searches. Even though it suffers from the lowered luminosity at LHCb, CODEX-b still shows some reach for the heavy \bivo regime, for lower squark mass, $M_{\rm sq}\sim 1$ TeV and heavy bi$\nu$o, $\Mbivo > 600$ GeV. It is also worth noting that missing energy searches can only put a lower limit on $\Lambda_M$ while MATHUSLA and CODEX-b can target specific values of both \bivo mass and the messenger scale.

Having relied on the clean and clear LLP analysis prescription described in \cite{Dercks:2018eua}, we also briefly describe the differences of the two works. In \cite{Dercks:2018eua} the authors study an $R$-parity-violating MSSM scenario with the term  $\lambda_{ijk}^\prime L_iQ_jD^c_k$ and look at a scenario where the lightest neutralino is produced in meson decays and decays into lighter mesons and charged leptons via this term. In the model we study, the $\lambda'$ term is not allowed due to the $U(1)_{R-L}$ symmetry. It is expected to be generated when $U(1)_{R-L}$ is broken, but its size is suppressed by $m_{3/2}/\Lambda_M$ compared to the $U(1)_{R-L}-$symmetric terms we investigate. 

%%%%%%%%%%%%%%%%%%%%%%%%%%%
\section{Summary $\boldsymbol{\&}$ Conclusions}
\label{sec:summary}
%%%%%%%%%%%%%%%%%%%%%%%%%%%

In this work we studied a model which presents an LLP, namely the bi$\nu$o, which is directly related to the  explanation of neutrino masses. In \cite{Coloma:2016vod} it was shown that in a 
 $U(1)_{R-L}$--symmetric supersymmetric model the observed  neutrino mass spectrum can be achieved via an inverse seesaw mechanism, in which the role of right-handed neutrinos is played by  the pseudo-Dirac bi$\nu$o. The lifetime of the \bivo depends on its mass and the messenger scale. The smaller the \bivo  mass and/or the larger the messenger scale the longer lived is the bi$\nu$o.
 Hence probing the \bivo lifetime in this scenario can lead to insights not only on the SUSY spectrum but also on SUSY breaking. If the \bivo is long-lived, it escapes ATLAS or CMS as missing energy and depending on the \bivo mass, \bivo momentum and the messenger scale, it potentially decays inside dedicated LLP experiments like FASER, CODEX-b, or MATHUSLA. 
 
 In order to study the potential of the LLP experiments, we first recast the most recent jets+$\slashed{E}_T$ searches from ATLAS, using $\sqrt{s}=13$ TeV and $\mathcal{L}=36~\text{fb}^{-1}$ to obtain the currently allowed parameter space. Then we simulated the reach of these experiment assuming a luminosity of $\mathcal{L}=3~\text{ab}^{-1}$. Our results are summarized in \Cref{fig:LLPexc}. We considered two regimes for the \bivo mass. \textbf{(i)} A light bi$\nu$o, with mass below the $W$ boson mass, decays through off-shell weak gauge or the Higgs bosons to a 3-body final state. In this case the current constraints from jets+$\slashed{E}_T$ searches exclude squark masses below 1.4 TeV independent of the \bivo mass for $\Lambda_M\gtrsim O(10~{\rm TeV})$. We then look at the reach of dedicated LLP detectors for this mass regime for $\Lambda_M>100$ TeV. We see that FASER and CODEX-b are not competitive in this scenario, both of which can only cover a low squark-mass region which is already excluded by jets+$\slashed{E}_T$ searches. However MATHUSLA is sensitive to messenger scales between $100-10^5$ TeV, part of which will not be covered by jets+$\slashed{E}_T$ searches at $\mathcal{L}=3~{\rm ab}^{-1}$. \textbf{(ii)} For heavy bi$\nu$os with mass above the Higgs mass the decay channel is into on-shell weak gauge bosons or the Higgs boson. In this regime the decay width is larger and current jets+$\slashed{E}_T$ constraints exclude \bivo masses below 500 GeV for squark masses below 1.3 TeV. In this regime MATHUSLA again has the best reach amongst the proposed LLP detectors. It can probe \bivo  and squark masses  up to 2 TeV for messenger scales between $10^9-10^{11}$ TeV. CODEX-b also becomes more competitive compared to the low bi$\nu$o-mass regime. 

Another possibility to search for a long-lived \bivo in this model is to search for displaced vertices in ATLAS or CMS. Recent analysis which rely on prompt jets and a displaced dilepton vertex \cite{Aad:2019tcc} have shown that  such analyses are possible with low background rates. The signature in our model scenario are two prompt jets and a displaced vertex with a combination of jets, charged leptons and missing energy. One significant difference between our model and generic RPV MSSM models this analysis relied on is that the charged leptons in the \bivo decays does not point to a vertex, see, \emph{e.g.}, \Cref{fig:LLPsignal}. In the absence of a public analysis, we will leave the study of displaced vertices inside LHC detectors in our model for future work.

In our previous analysis on short-lived bi$\nu$os \cite{Fox:2019ube},  we focused on a messenger scale of 100 TeV and \bivo masses between 100 GeV-1 TeV. In this current work we extended that parameter region to a bi$\nu$o mass of 1 GeV as well as much higher messenger scales, up to $10^{11}$ TeV. Together, we have analyzed a vast region of parameter space in this model.

\begin{acknowledgments}
The authors cordially thank Pilar Coloma for collaboration in the early stages of this work and Paddy Fox for enlightening conversations and a careful reading of a draft of this paper. The authors also welcome the young Roberto Jr. Coloma to the particle physics world. JG acknowledges support from the US Department of Energy under Grant Contract DE-SC0012704. SI is supported by the NSF via grant number PHY-191505.
\end{acknowledgments}

\bibliographystyle{JHEP}
\bibliography{ref}{}

\end{document}